# Application of the chemical master equation and its analytical solution to the illness-death model

Ralph Brinks[1,2]

1) Chair for Medical Biometry and Epidemiology, Witten/Herdecke University, Faculty of Health/School of Medicine, D-58448 Witten, Germany

2) Institute for Biometry and Epidemiology, German Diabetes Center, D-40225 Düsseldorf, Germany

Corresponding author:
E-Mail: ralph.brinks@uni-wh.de

## Abstract

The aim of this article is relating the chemical master equation (CME) to the illness-death model for chronic diseases. We show that a recently developed differential equation for the prevalence directly follows from the CME. As an application, we use the theory of the CME in a simulation study about diabetes in Germany from a previous publication. We find a good agreement between the theory and the simulations.

Key words: prevalence, chronic diseases, diabetes, epidemiology, monomolecular reaction.

# Introduction

In 2007, Jahnke and Huisinga presented an analytical solution to the chemical master equation (CME) for stochastic modeling of biochemical reaction systems [Jah07]. The authors confine their theory to monomolecular reactions, i.e., reactions where a substance $S_j$ converts to another substance $S_k$, written as $S_j \rightarrow S_k$, or the substance $S_j$ flows in or out from a source (* $\rightarrow S_j$) or to a sink ($S_j \rightarrow$ *). These are the types of reactions that can be transferred to the illness-death model in epidemiology. In the illness-death model, a subject changes its state from *non-diseased* to *diseased* (or vice versa) or changes its state to *dead* (from either state *non-diseased* or *diseased*).

The aim of this article is applying the theory of [Jah07] for solving the CME to the illness-death model. We show that a recently developed differential equation for the prevalence of the disease considered in the illness-death model [Bri13] directly follows from the CME. As an application, we use the theory of [Jah07] in a problem about a chronic condition from a previous publication [Bri18].

# Illness-death model and the chemical master equation

We consider the illness-death model as depicted in Figure 1. At start (time $t = 0$), each subject of the population under consideration is assigned to exactly one health state: *non-diseased* (state $S_1$), *diseased* (with respect to a specific condition, state $S_2$) and *dead* (state $S_3$). Subjects may change their state according to the arrows in Figure 1 as time $t$ evolves. The non-negative transition rates are denoted by $c_{jk}$ and usually depend on time $t$, i.e., $c_{jk} = c_{jk}(t)$. The transitions $S_3 \rightarrow S_1$, and $S_3 \rightarrow S_2$ are not possible, which means that the rates $c_{31}$ and $c_{32}$ are assumed to be zero for all times $t$. Furthermore, for ease of notation we define $c_{jj} = 0$.

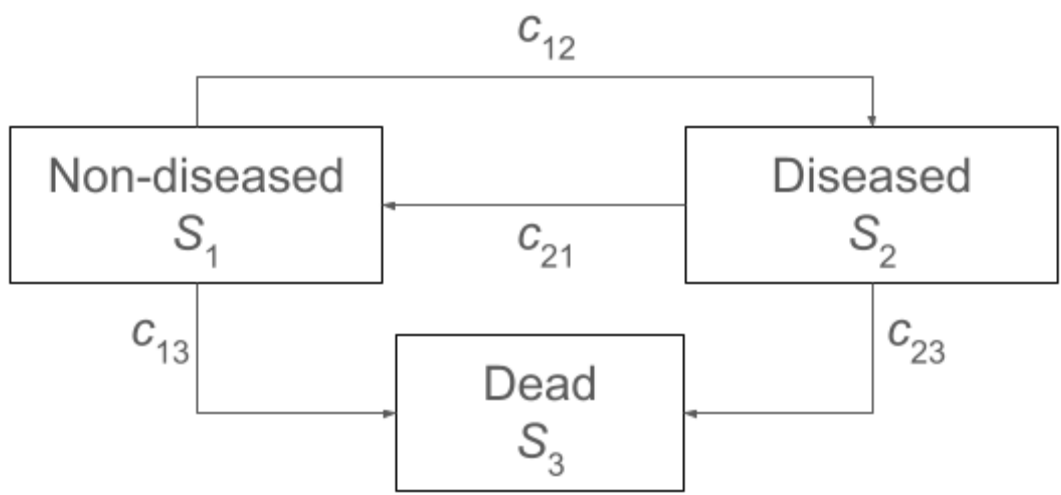

Figure 1: Illness-death model with transition rates $c_{jk}$ between the states $S_j$.

In terms of [Jah07] the health states $S_j$ in Figure 1 are the types of substances. Jahnke and Huisinga consider monomolecular reactions as described in the introduction above. For our purpose we assume that the population has the size $N$, where $N$ is a positive integer, $N \in \mathbb{N} = \{1, 2, \ldots\}$, and at $t = 0$ all subjects are in the non-diseased state $S_1$. Let $X_j = X_j(t)$ be the non-negative integer number of subjects in state $S_j$, $j = 1, 2, 3$, at time $t \geq 0$. Then, we have $X_1(0) = N$ and $X_2(0) = X_3(0) = 0$. For ease of notation, we define $X(t) = (X_1(t), X_2(t), X_3(t))$.

Again, applying the terminology of [Jah07] to our situation, for $x \in \mathbb{N}_0^3 = \{0, 1, 2, \ldots\}^3$, we consider the probability

$$\mathbb{P}(t, x) = \mathbb{P}(X(t) = x) = \mathbb{P}(X_1(t) = x_1, X_2(t) = x_2, X_3(t) = x_3).$$

Then, the chemical master equation (CME) reads as

$$\partial_t \mathbb{P}(t, x) = \sum_{j,k=1}^{3} c_{jk}(t) \left[ (x_j + 1)\, \mathbb{P}(t, x + \varepsilon_j - \varepsilon_k) - x_j\, \mathbb{P}(t, x) \right],$$

where the $c_{jk}$ are the transition rates (see Figure 1) and $\varepsilon_j$ is the $j$-the unit vector, $j = 1, 2, 3$, (e.g., $\varepsilon_3 = (0, 0, 1)$). This form of the CME is a special case of Eq. (8) in [Jah07].

The most important finding of [Jah07] for the illness-death model is Proposition 1 in [Jah07]. The proposition states that if the initial condition at $t = 0$ is the multinomial distribution $\mathbb{P}(0, x) = \mathcal{M}(x, N, p_0)$ for a parameter vector $p_0 \in [0,1]^3$, then for $t > 0$ it holds $\mathbb{P}(t, x) = \mathcal{M}(x, N, p(t))$ where the parameter $p(t)$ is the solution of a system of ordinary differential equations (ODEs):

$$\frac{d}{dt} p(t) = p'(t) = A(t)\, p(t) \tag{1}$$

with initial condition $p(0) = p_0$. The matrix $A(t)$ in Eq. (1) is given by $A = (a_{jk})$ defined via $a_{jk} = c_{kj}$, for $j, k = 1, 2, 3$ with $j \neq k$ and $a_{kk} = -(c_{k1} + c_{k2} + c_{k3})$, $k = 1, 2, 3$ (see Eq. (5) in [Jah07]). Note that we defined $c_{kk} = 0$.

The definition of the matrix $A$ for the illness-death model shown in Figure 1 is

$$A(t) = \begin{bmatrix} -c_{12}(t) - c_{13}(t) & c_{21}(t) & 0 \\ c_{12}(t) & -c_{21}(t) - c_{23}(t) & 0 \\ c_{13}(t) & c_{23}(t) & 0 \end{bmatrix}$$

Using $p(t) = (p_1(t), p_2(t), p_3(t))$, we obtain following linear system of ODEs:

$$\begin{align}
p_1' &= -(c_{12} + c_{13})\, p_1 + c_{21}\, p_2 \tag{2a} \\
p_2' &= c_{12}\, p_1 - (c_{21} + c_{23})\, p_2 \tag{2b} \\
p_3' &= c_{13}\, p_1 + c_{23}\, p_2. \tag{2c}
\end{align}$$

Note that $p_0 \in [0,1]^3$ is three-dimensional, but $p_1, p_2, p_3 \in [0,1]$ are scalars.

If we define $\pi(t) := p_2(t)/[p_1(t) + p_2(t)]$ for $p_1(t) + p_2(t) > 0$ and insert Eqs. (2a) and (2b) into $\pi' = [p_1 p_2' + p_1' p_2]/[p_1 + p_2]^2 = [(1 - \pi) p_2' - \pi p_1']/[p_1 + p_2]$, we obtain

$$\pi' = (1 - \pi)\, \frac{c_{12} p_1 - (c_{12} + c_{23}) p_2}{p_1 + p_2} + \pi\, \frac{(c_{12} + c_{13}) p_1 - c_{21} p_2}{p_1 + p_2},$$

and by algebraic formulations we find

$$\pi' = (1 - \pi)\left(c_{12} - \pi (c_{23} - c_{13})\right) - \pi c_{21}. \qquad (3)$$

Equation (3) is a scalar ODE of Riccati type. If $X(t) = (X_1(t), X_2(t), X_3(t))$ is multinomially $\mathcal{M}(x, N, p(t))$ distributed, we have the expectation $\mathbb{E}(X(t)) = N\, p(t)$ and hence $\pi = p_2/[p_1 + p_2] = \mathbb{E}(X_2)/[\mathbb{E}(X_1) + \mathbb{E}(X_2)]$, which implies that $\pi$ is the prevalence of the expected (average) numbers of subjects in the associated states. In this sense, Equation (3) is equivalent to Equation (3) in [Bri15], which was obtained using a different proof. For the special case $c_{21} = 0$, Equation (3) above is Equation (9) of [Bri18].

## Simulation

To apply the theory presented in the previous section, we choose a test example from [Bri18] motivated by the situation of type 2 diabetes in Germany. Diabetes is assumed to be irreversible, i.e., the remission rate $c_{21}$ vanishes, $c_{21}(t) = 0$ for all times $t \in [0,\infty)$. The incidence rate $c_{12}$ is given by $c_{12}(t) = \max(0, t - 30)/2000$ and the mortality rates $c_{13}$ and $c_{23}$ are assumed to be of Gompertz-type $c_{13}(t) = \exp(-10.7 + 0.1\, t)$ and $c_{23}(t) = \exp(-10 + 0.1\, t)$. As initial condition for ODE (1), we choose $p_0 = (1, 0, 0)$. Then, the initial value problem is solved by the classical Runge-Kutta method of fourth order (`rk4` in the R package `deSolve` [Soe10]). The result for the components $p_1$, $p_2$, and $p_3$ of $p = p(t)$ is shown in Figure 2.

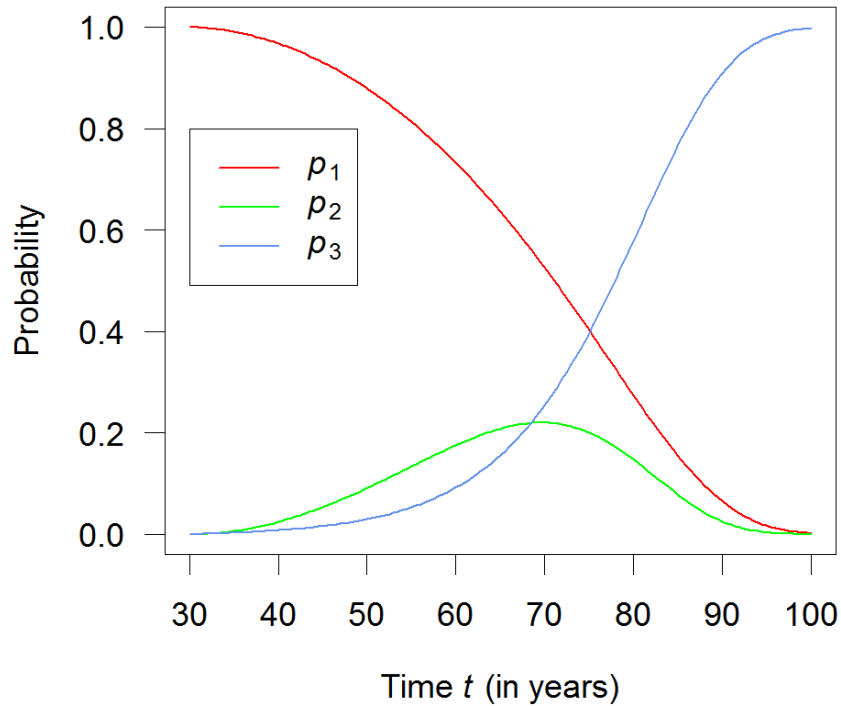

**Figure 2: Components $p_1$, $p_2$, and $p_3$ of the solution $p = p(t)$ of ODE system (1) with initial condition $p_0 = p(0) = (1, 0, 0)$.**

In order to demonstrate that the solution $p = p(t)$ of ODE system (1) with initial condition $p_0 = p(0) = (1, 0, 0)$ can be used to characterize the expected values $(\mathbb{E}(X_1), \mathbb{E}(X_2), \mathbb{E}(X_3)) = N\,p$, we calculate 100 realizations of $\mathcal{M}(x, N, p_0)$ random variables with $N = 200$, $p_0 = (1, 0, 0)$ and $x = (200, 0, 0)$ undergoing transitions according to the rates $c_{jk}$ described above. The associated curves $N\,p(t)$ where the components of $p(t)$ are shown in Figure 2 are then compared to the 100 realizations. Simulations of the random variable $X = (X_1, X_2, X_3)$ have been done with the algorithm described in [Bri14]. For better presentation only values in 10-years steps at $t = 30, 40, \ldots, 90$ years are presented. The results are shown in Figure 3.

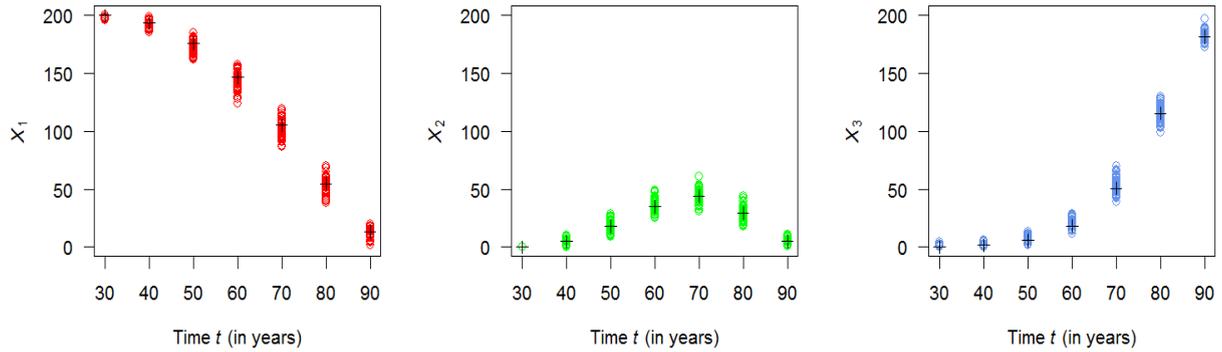

**Figure 3: 100 realizations of multinomially distributed random variables $X = (X_1, X_2, X_3)$ ($X_1$: red circles (left panel), $X_2$: green circles (middle) and $X_3$: blue circles (right panel) and associated expectations $\mathbb{E}(X(t)) = N\,p(t)$ (black crosses).**

Finally, we assess if the prevalence $X_2/[X_1 + X_2]$ computed by the components of the random variable $X = (X_1, X_2, X_3)$ in the 100 realizations agree with the solution $\pi$ of ODE (3) with initial condition $\pi_0 = 0$. Again, the associated initial value problem has been solved with the Runge-Kutta method of fourth order (as above). The results are shown in Figure 4 - for better comparison again at $t = 30, 40, \ldots, 90$ years.

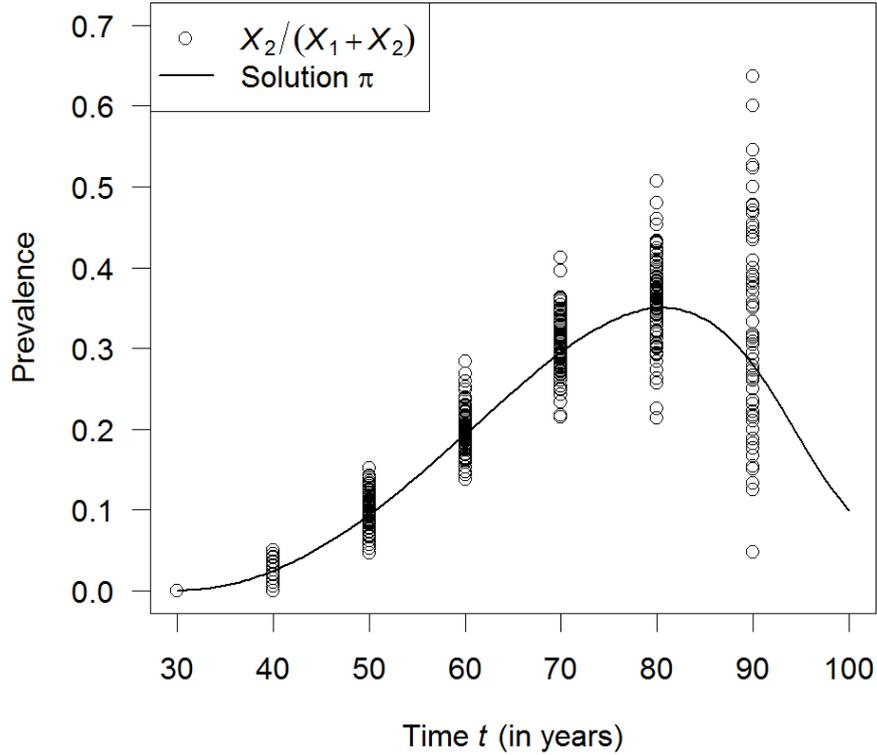

**Figure 4:** Prevalence $X_2/[X_1 + X_2]$ in 100 realizations of multinomially distributed random variables $X = (X_1, X_2, X_3)$ (black circles) compared to the solution $\pi$ of ODE (3) with initial condition $\pi_0 = 0$ (solid line).

## Summary

We could show that the theory of the chemical master equation (CME) can be used to obtain theoretical insights about the frequently used illness-death model. In case $X = (X_1, X_2, X_3)$ is multinomially distributed, applying the CME yields that the (average) prevalence $\pi = \mathbb{E}(X_2) / [\mathbb{E}(X_1) + \mathbb{E}(X_2)]$ is a solution of the Riccati ODE (3), which has been proven elsewhere without use of the CME. The advantage of the proof using the CME is that the three-dimensional random variable $X$ has components that are integers. Moreover, interpretation of $\pi$ in terms of random variables allows the use of inference and secondary order statistics. For instance, for multinomially distributed $X = (X_1, X_2, X_3)$ and for $j, k = 1, 2, 3$ with $j \neq k$, we have $\mathrm{Cov}(X_j, X_k) = -N p_j p_k$ and $\mathrm{Var}(X_k) = N p_k (1-p_k)$.

Apart from the theoretical findings, applicability has been demonstrated in a simulation study motivated by diabetes in Germany. One hundred realizations of the random variable $X(t)$ are compared with the curves $t \mapsto N p(t)$ where $p(t)$ is the solution of ODE system (1). The realizations of $X$ have been simulated with a previously described discrete event simulation especially designed for the illness-death model [Bri14]. By visual inspection, we find a good agreement between the theoretical findings and the simulations.